\documentclass[a4paper]{article}
\usepackage{Odyssey2022}
\usepackage{epsfig,amssymb,amsmath}
\usepackage{xcolor}

\newcommand{\norm}[1]{\left\lVert#1\right\rVert}
\usepackage[framemethod=tikz]{mdframed}
\usepackage{tikz} 
\usepackage{graphicx}
\usepackage[colorinlistoftodos]{todonotes}
\usepackage[colorlinks=true, allcolors=blue]{hyperref}
\usepackage{enumitem}
\usepackage{verbatim} 
\usepackage{subcaption}
\usepackage{booktabs}
\usepackage{multirow}

\newcommand{\killtext}[1]{}
\def\Rset{\mathcal{R}}
\def\rvec{\mathbf{r}}
\def\lvec{\mathbf{l}}
\def\Kmat{\mathbf{K}}
\def\Mmat{\mathbf{M}}
\ninept

\title{Analyzing speaker verification embedding extractors and back-ends under language and channel mismatch}

\name{\begin{tabular}{c} Anna Silnova$^{1}$,
Themos Stafylakis$^{2}$,
Ladislav Mo\v{s}ner$^{1}$,
Old\v{r}ich Plchot$^{1}$,\\
Johan Rohdin$^{1}$,
Pavel Mat\v{e}jka$^{1}$,
Luk\'{a}\v{s} Burget$^{1}$,
Ond\v{r}ej Glembek$^{1}$,
Niko Brummer${^{3}}$
\end{tabular}}

\address{$^{1}$Brno University of Technology, Speech@FIT and IT4I Center of Excellence, Brno, Czechia\\
$^{2}$Omilia - Conversational Intelligence, Athens, Greece \\
$^{3}$Phonexia, South Africa\\}

\begin{document}
\maketitle
\begin{abstract}
In this paper, we analyze the behavior and performance of speaker embeddings and the back-end scoring model under domain and language mismatch.  We present our findings regarding ResNet-based speaker embedding architectures and show that reduced temporal stride yields improved performance. We then consider a PLDA back-end and show how a combination of small speaker subspace, language-dependent PLDA mixture, and nuisance-attribute projection can have a drastic impact on the performance of the system. Besides, we present an efficient way of scoring and fusing class posterior logit vectors recently shown to perform well for speaker verification task. The experiments are performed using the NIST SRE 2021 setup.  
\end{abstract}

\section{Introduction}
\killtext{
-As usual, SRE21 is focused on telephone data (most of the trials)

- trials (target and non-target) with enrollment and test segments originating from different source types (i.e., CTS and AfV)

- trials (target and non-target) with enrollment and test segments spoken in different languages (i.e., cross-lingual trials)

-We participated in a fixed condition (allows for simpler comparison between sites).

-We developed a narrowband system, utilizing the newly released NIST SRE CTS Superset, LDC2021E08 as well as the Voxceleb 1 and 2, which we downsampled.

}

In this paper, we will provide an analysis of an integral part of every state-of-the-art speaker recognition system, which is a DNN-based embedding extractor and a subsequent speaker back-end that provides probabilistic verification scores for individual trials. We focus on the most recent NIST SRE 2021 which provided us with an opportunity for experimentation and this analysis.

In speaker recognition (SR), deep neural networks (DNNs) have nowadays firmly established their role (or roles) in state-of-the-art systems. The past line of research focused on 
modeling the fixed-length utterance representations, such as i-vectors~\cite{DehakN_TASLP:2010} obtained as maximum a-posteriori estimates of a latent variable in generative factor analysis model. The obtained i-vector representations (generative embeddings) were subsequently modeled by probabilistic linear discriminant analysis (PLDA)~\cite{prince:iccv:2007}, a technique introduced in face verification.

DNNs have been gradually incorporated into the speaker recognition pipeline, through replacing or improving one or more of the components of an i-vector + PLDA system (e.g. feature extraction, calculation of sufficient statistics, i-vector extraction or PLDA classifier). 
For instance on the front-end level, 
  employment of DNN bottleneck features (BNF) instead of conventional MFCC features~\cite{lozano_odyssey_2016}, or simply concatenating BNF and MFCCs~\cite{Matejka:ICASSP:2016} and more recently learning the feature extraction directly from the raw audio~\cite{jung2019rawnet,jung2020improved,lin2020wav2spk}, was proposed. Later in the modeling stage, NN acoustic models were proposed 
to replace generative Gaussian mixture models (GMM) for extraction of sufficient statistics~\cite{Lei_icassp_2014,kennydeep}, or for complementing or substituting the PLDA~\cite{Novoselov_interspeech_2015,Bhattacharya_SLT16}, \cite{Ghahabi_icassp_2014}, respectively.

New deep learning works have 
logically resulted in attempts to train a larger DNN directly for speaker recognition tasks, i.e., binary classification of two utterances as a \emph{target} or a \emph{non-target} trial~\cite{heighold_icassp_2016, zhang_slt_2016, sreEndEnd:Snyder2016, rohdin:icassp:2018}. Such systems are denoted as \emph{end-to-end} systems and were proven competitive for text-dependent tasks~\cite{heighold_icassp_2016, zhang_slt_2016, bhattacharya2016deep} as well as for text-independent tasks considering short test utterances and an abundance of training data~\cite{sreEndEnd:Snyder2016}. On text-independent tasks with longer utterances and moderate amount of training data, the i-vector motivated end-to-end system~\cite{rohdin:icassp:2018} outperformed generative baselines, but at the cost of high complexity
during training.

While the fully end-to-end SR systems have been struggling with large requirements on the amount of training data (often not available to the researchers) and high computational costs, the focus on speaker recognition has partially shifted back to generative modeling, but now with utterance representations obtained from a single DNN. Such DNN takes the frame-level features of an utterance as an input and directly produces an utterance-level representation, usually referred to as an \emph{embedding}~\cite{Variani_icassp_2014, heighold_icassp_2016, zhang_slt_2016, Bhattacharaya_interspeech_2017, xvec:Snyder2016,stafylakis2019self}. The embedding is obtained by the means of a \emph{pooling mechanism} (for example taking the mean) over the frame-wise outputs of one or more layers in the~DNN~\cite{Variani_icassp_2014}, or by the use of a recurrent DNN~\cite{heighold_icassp_2016}. 
One effective approach is to train the DNN for classifying a set of training speakers, i.e., using multi-class training~\cite{Variani_icassp_2014, Bhattacharaya_interspeech_2017,xvec:Snyder2016}. In order to perform speaker verification, the embeddings are extracted and used in a standard back-end, e.g. PLDA ~\cite{Bhattacharaya_interspeech_2017, xvec:Snyder2016}.

Most recently, deep convolutional neural networks (DCNNs) models such as ResNet~\cite{he2016deep} play the core role in the SR systems and are continuously replacing feed forward DNNs~\cite{xvec:Snyder2016} for embedding extraction. DCNNs are often fine-tuned via optimizing angular margin loss as in face verification. This approach offered the best results in the domain of English wide-band data~\cite{Zeinali_VOXCELEB_2019} and has been competitive in more challenging and less data-rich domain of telephone Tunisian Arabic data~\cite{ABC_SRE_CTS_2019} as well as in the most systems submitted to the recent NIST SRE 2020 and 2021 challenges which always showcase the latest state-of-the-art modeling techniques. 

Great deal of our analysis will explore subtle changes in the standard ResNet architecture such as changing the temporal strides or different pooling mechanisms. We will compare the performance of operationally practical and relatively small ResNet34 with a much larger ResNet152 and we will provide an analysis with fine-tuning these models for longer duration segments often present during inference as these are typically in sharp contrast with very short (and same length) segments used during training.

As it is typical in NIST SREs, also the last SRE21 brought a challenging scenario of a channel, language, and data mismatch w.r.t. training data. We are therefore exploring 
an approach to incorporate language information into PLDA framework, various nuisance variability suppression techniques, and score normalization. These methods, in our experiments, led to better results than using a simple cosine-distance scoring. We also experiment with taking the embeddings from different parts of the DNN and use our back-ends to model directly the output of the pooling layer, standard (x-vector) embedding, and also we provide an insight into scoring with class posterior logits from the very end of the ResNet embedding extractor.

\section{Embedding extractors and Back-end model}
\label{sec:AudioSystems}

In this section, we present the basic architecture of the embedding extractors used in the experimental part of this paper. Also, we describe in detail the back-end approach used in the majority of experiments. When presenting the experiments we will always describe the differences in the setup with respect to the models and approaches described here.

\subsection{ResNet Architectures} 

Residual Networks (ResNets), together with their variants (e.g. ResNeXt, Res2Net \cite{zhou2021resnext}) are standard choices in speaker recognition research. In this paper, we examine two different standard ResNet architectures, namely ResNet34 and ResNet152. ResNet34 provides a good compromise between accuracy and computational efficiency, e.g. it can be deployed in CPU-based production systems. On the other hand, ResNet152 is a very deep architecture that can yield state-of-the-art results but requires GPU in runtime for being deployed in production systems that perform real-time processing. In terms of architectural investigation, we experiment with the following directions. 
\subsubsection{Statistics pooling}
Apart from the typical mean and standard deviation (std) statistics pooling, we examine using only standard deviation features in the statistics pooling layer. The approach was examined in \cite{wang2021revisiting} and appears to generalize better, at least in cases of dataset-shifts between training and test. Furthermore, we experimented with the recently proposed correlation pooling, which showed good improvements in VoxCeleb~\cite{stafylakis2021speaker}.

\subsubsection{Temporal stride} We also experiment with reducing the temporal stride of the ResNet blocks. The temporal stride per ResNet stage is set to [1,2,1,2] or [1,1,2,1] (i.e. a cumulative stride equal to 4 or 2 instead of the standard 8) while the frequency stride is the typically used [1,2,2,2]. The motivation is to reduce the receptive field in order to model shorter speech patterns, which should be more language-independent. We should mention though that reducing the stride results in increased computational requirements, and hence the model becomes less efficient. 

\subsubsection{Extracting alternative speaker representations} Another set of experiments is related to the layer from which the speaker representation is extracted. One alternative to the standard embeddings is extracting the statistics. Our motivation is that embeddings are susceptible to overfitting the training speakers and languages, as they interact directly with the classification head. Consider for example TDNN-based x-vectors, where two low-dimensional representations can serve as embeddings. The experiments clearly show that embeddings extracted from the input to the classification head are inferior~\cite{snyder2018Xvectors}. On the other hand, in ResNet-based architectures there is a single low-dimensional representation after statistics pooling. We experiment with modifying the architecture by adding a second hidden layer however the results were discouraging. We, therefore, consider to extract the statistics instead. Statistics allows us to experiment with unsupervised dimensionality reduction methods (e.g. PCA), use a training set that is more diverge compared to that of the extractor, and possibly retain directions that are more discriminative for languages and domains not included in the training set.

Apart from using the statistics, we examine the recently proposed method of extracting the logits, i.e. the projection of the embeddings to the classification head (prior to the softmax function). The method is proposed in ~\cite{stc_sysdescr} and yields state-of-the-art results in SRE21. We moreover show that the method can be implemented more efficiently using Cholesky (or other equivalent) decomposition of symmetric positive (semi-)definite matrices. 

\subsubsection{Fine-tuning on long durations} Several papers have reported improvements by fine-tuning the extractor on long durations for the last few epochs. However, most of the results are based on VoxCeleb and use cosine similarity scoring. It is therefore interesting to examine whether similar improvements will be attained on more challenging setups and with a PLDA back-end.

\subsection{Back-end Architectures}%
\label{sec:backend_theory}
In most of the experiments described in this paper, we follow the same approach to pre-processing the embeddings and training the back-end:

\subsubsection{Nuisance attribute projection}
We start by removing from the data the direction corresponding to speaker gender by nuisance attribute projection (NAP)~\cite{solomonoff2005,aronowitz2014inter}. Further on in Section~\ref{sec:nap} we present the experiment motivating for suppressing gender information. Then, we proceed with centering the data, LDA reducing dimensionality of the embeddings to 100 (see Section~\ref{sec:backend} for details), and length normalization. 

\subsubsection{Mixture of language-dependent PLDAs}
After data pre-processing, we train a mixture of 3 PLDA models~\cite{senoussaoui2011mixture,garcia2012multicondition}: each component of the mixture is a PLDA trained on the data coming from one of three languages: English, Cantonese, and Mandarin. At test time, we score each trial with each language-dependent PLDA model, the final score of the mixture is a weighted average of the scores of individual mixture components:
\begin{align}
    s_{e,t}=\sum_{p\in \{eng,cmn,yue\}}w^p_{e,t}s^p_{e,t},
\end{align}
where $s^p_{e,t}$ are LLR scores computed with one of the mixture components (PLDA model $M_p$) and $w^p_{e,t}$ are corresponding weights given by:
\begin{align}
    w^p_{e,t}=\frac{1}{2} \left(P(\Rset_{e}\mid M_p)+P(\Rset_{t}\mid M_p)\right).
\end{align}
In the expression above, $\Rset_e$ and $\Rset_t$ are enrollment and test embeddings (one or three in case of multi-session enrollment models). Likelihoods $P(\Rset\mid M_p)$ can be estimated using e.g. equation (19) of~\cite{BrummerSimPLDA}.

Notice, that our approach to estimating the score is somewhat simpler than that of~\cite{senoussaoui2011mixture}, where the correct LLR of the mixture was estimated. At early stage of the development, we compared these two approaches but observed a slight performance degradation when using more complicated by-the-book scoring with the mixture of PLDA models and decided to proceed with the easier approach described above.

\section{Experiments and results}
\subsection{Experimental setup}
\label{HXV:data}
For training our embedding extractors we used three databases, namely NIST CTS Superset~\cite{sre_cts_superset}, Voxceleb 1 and 2~\cite{Nagrani19,chung2018voxceleb2}. In all our experiments, the systems are trained on 8kHz data, all 16kHz data from the aforementioned datasets are downsampled to 8kHz. In total, there are 14096 training speakers. We used Kaldi style augmentation with MUSAN database~\cite{snyder2015musan}. At the input of the embedding extractors there are 64 mel filter banks with frequency band limited to 20-3800Hz.

When training the back-end model, we use only NIST CTS Superset (English, Cantonese, and Mandarin recordings from this set). We use the embeddings extracted from the original recordings along with one augmented copy of the data. Each recording is augmented with one augmentation that was randomly selected out of five types of augmentation used when training the extractor. There are approximately 7000 speakers used for training the back-end. 

In all of our experiments, we test the performance on NIST SRE 21 evaluation set. The particulars of the evaluation can be found in the evaluation plan~\cite{Sadjadi:2021}. The relevant details for our system design choices are: there are no cross-gender trials; the majority of the utterances are in English, Cantonese, and Mandarin; there are cross-language trials.  We report the results in terms of Equal Error Rate (EER) and minimum cost (min\_C) as computed by the scoring tool released by NIST as a part of the evaluation. 

Below, we present the results of the experiments where we vary different aspects of the embedding extractors described in Sections~\ref{sec:resnet34} and~\ref{sec:resnet152} and the back-end model of Section~\ref{sec:backend_theory}. We start with the experiments on embedding extractors and then continue with the back-end related experiments.
\subsection{Effect of temporal stride}
\label{sec:stride}
As we know, the evaluation set mostly consists of non-English data, while the majority of the recordings used for training the embedding extractors are in English. Hence, there is a need to compensate for the language mismatch between training and evaluation data. Our assumption is that by reducing the temporal context, we encourage the network to train on more language-independent patterns. To verify this assumption, we trained several ResNet networks with three settings for temporal stride: strides were set to [1,2,2,2], [1,2,1,2], or [1,1,2,1] per ResNet stage (i.e. a cumulative stride equals to 8, 4 and 2, respectively), frequency stride in all cases was fixed to [1,2,2,2]. We experimented with two different pooling mechanisms (more information is given in Section~\ref{sec:pooling}). The back-end model for all experiments was as described in~\ref{sec:backend_theory}.

The results of the extractors with varying temporal strides are shown in Table~\ref{tab:results:stride}. As one might notice, reducing the cumulative temporal stride from 8 to 4 (compare lines 1 and 2 of the table) results in a noticeable performance improvement for both pooling mechanisms in case of ResNet34. For the larger embedding extractor, the performance improvement is not as pronounced but still non-negligible. However, reducing the stride even further degrades the performance (we have the results only for ResNet34 with std pooling) indicating that for successful performance the network has to observe at least some temporal dependencies in the data.  

\begin{table*}[h]
  \caption{
  Comparison of speaker verification performance of ResNet embedding extractors with different temporal strides (two different pooling approaches for ResNet34). The back-end used is a mixture of 3 language-dependent PLDA models.} 
  \label{tab:results:stride}
  \centering
   \begin{tabular}{ c l  c c  c c c c }

    \toprule
     & \multirow{3}{*}{\textbf{Temp stride}} & \multicolumn{4}{c}{\textbf{ResNet34}} & \multicolumn{2}{c}{\textbf{ResNet152}}\\
    & &\multicolumn{2}{c}{\textbf{std}} &\multicolumn{2}{c}{\textbf{corr}}&&  \\
     & & \textbf{min\_C}  & \textbf{EER (\%)} & \textbf{min\_C} & \textbf{EER (\%)}& \textbf{min\_C} & \textbf{EER (\%)} \\
    \midrule
1 &8:[1,2,2,2] & 0.495 & 8.66 & 0.553& 9.84&	0.412& 6.32 \\ 
    2 & 4:[1,2,1,2] & 0.473 & 8.38 & 0.520 & 9.44&	0.407&6.28 \\ 
    3  &2:[1,1,2,1] & 0.504 & 9.07 & -&-&-&-\\ 
    \bottomrule
  \end{tabular}
  \vspace{-4mm}
\end{table*}
\subsection{Statistics pooling}
\label{sec:pooling}
Here, we compare three approaches to pool frame-level representations in the embedding extractor. The first option we consider is the standard way of statistics pooling when both mean and std are computed. Second, we consider the case when only standard deviation features are used for statistics pooling. In~\cite{wang2021revisiting}, this approach was shown to generalize better when there is a mismatch between training and test data. Finally, we use the approach of~\cite{stafylakis2021speaker} where correlation pooling is used instead of mean and std.

We compare these three approaches by training ResNet34 embedding extractors with the aforementioned pooling mechanisms. The results are presented in Table~\ref{tab:results:pooling}.
As can be seen, both mean+std and std alone result in a similarly good performance in terms of min\_C, however, in terms of EER, using just standard deviation brings considerable gain. Regarding correlation pooling, we notice that there is a notable performance degradation compared to the other two options.
\begin{table}[h]
  \caption{
  Comparison of performance of embedding extractors utilizing different pooling strategies. In all cases, the architecture we use is ResNet34 with cumulative temporal stride 4. The back-end used is a mixture of 3 language-dependent PLDA models.} 
  \label{tab:results:pooling}
  \centering
   \begin{tabular}{ c l  c c }

    \toprule
     & \textbf{Pooling} & 
       \textbf{min\_C}  & \textbf{EER (\%)}  \\
    \midrule
1 &mean+std & 0.474 & 8.74 \\ 
    2 & std & 0.473 & 8.38  \\ 
    3  &correlation & 0.520 & 9.44 \\ 
    \bottomrule
  \end{tabular}
  \vspace{-4mm}
\end{table}
\subsection{Extracting statistics as speaker representations}
Typically, speaker embeddings are the activations of the low-dimensional penultimate layer of the extractor network. Hence, the embeddings extracted in this way might be overtuned towards training speakers and domains. To overcome this potential problem, we try to extract the speaker representation directly as the output of the statistics pooling layer. As this layer is further away from the classification head and has a higher dimension, our expectation is that such a representation would be less susceptible to overtraining to the speakers that were used when training the network. However, because such representations have relatively high dimensionality (2048 in our case), it is problematic to directly use them for training the back-end model. For this reason, we need to perform dimensionality reduction of  high-dimensional statistics first. We believe that if dimensionality reduction is trained on the set of speakers other than that was used for training the network then, even though the resulting embeddings are low-dimensional vectors, they will be more suitable for a general speaker recognition problems than the original embeddings.

In our experiments, we tried the simplest option for dimensionality reduction: Principal Component Analysis (PCA). PCA projection matrix is estimated on the back-end training set and reduces the dimensionality of statistics from 2048 to 256 (size of the original embeddings).

\subsection{Extracting logits as speaker representations}
In the opposite direction, the authors in~\cite{stc_sysdescr} propose to use high-dimensional vectors of class posterior logits (cl-embeddings) in place of conventional embeddings. They show performance improvements when such vectors are used in combination with cosine distance scoring. However, we want to point out that using high-dimensional vectors of logits is not necessary. Instead, one can compute a low-dimensional projection of the original embeddings such that scoring high-dimensional cl-embeddings will be equivalent to scoring low-dimensional projected embeddings. Below, we demonstrate it for cosine scoring, but a similar approach can be used for training a PLDA model.

When scoring cl-embeddings, we make use of the fact that the logits $\lvec$ are just high-dimensional projection of the original embeddings $\rvec$:
\begin{align}
    \lvec=\Kmat\rvec,
\end{align}
where $\Kmat$ is the projection matrix of shape $N_s \times d$ (number of training speakers and embedding dimension, respectively). Then the cosine score between enrolment and test becomes:
\begin{align}
    s_{e,t}=\frac{\lvec_e'\lvec_t}{\norm{\lvec_e}\norm{\lvec_t}}=\frac{\rvec_e'\Kmat'\Kmat\rvec_t}{\norm{\Kmat\rvec_e}\norm{\Kmat\rvec_t}}.
\end{align}
The matrix $\Kmat'\Kmat$ is a symmetric positive (semi-)definite matrix and can be considered as $N_s$ times the between-speaker covariance, estimated with the speakers contained in the training set of the extractor (recall that when AAM loss is used, the rows of $\Kmat$ have unit norm). By performing Cholesky decomposition we obtain the upper triangular $d\times d$ matrix $\Mmat$ such that $\Kmat'\Kmat=\Mmat'\Mmat$. As a result:
\begin{align}
s_{e,t}=\frac{\rvec_e'\Mmat'\Mmat\rvec_t}{\norm{\Mmat\rvec_e}\norm{\Mmat\rvec_t}}.
\end{align}
Thus, instead of extracting and scoring with high-dimensional logit vectors $\lvec=\Kmat\rvec$, we may equivalently extract low-dimensional vectors $\Mmat\rvec$.
Finally, an efficient implementation of the fusion method of logits (proposed also in~\cite{stc_sysdescr}) is derived in Appendix \ref{logit_fusion}.

We perform the experiment where we compare three approaches: conventional embeddings (line 1 of Table~\ref{tab:results:embd_stats}), outputs of the statistics pooling layer in combination with PCA (line 2), and embeddings projected with low-dimensional square matrix $\Mmat$, computed as described above (line 3). For all three types of speaker embeddings, we use either cosine scoring or train the mixture of language-dependent PLDA models as described in Section~\ref{sec:backend_theory}. In the latter case, $\rvec$ are projected onto $\Mmat$ and length-normalized, prior to LDA and a new length-normalization. Here, we perform the experiments only with ResNet34 embedding extractor with cumulative temporal stride set to 4 and standard deviation pooling. The results of this experiment are presented in Table~\ref{tab:results:embd_stats}. 

In case of cosine scoring, we observe major performance gains from using cl-embeddings compared to conventional embedding vectors, which agrees with the findings of~\cite{stc_sysdescr}. Using statistics in this scenario, however, results in severe degradation. For the second back-end approach, we do not observe such high variability in the results. Statistics, though, still provide inferior results compared to the other two types of speaker representations. One of the possible explanations is the dimensionality reduction approach that we use. In future, we plan to investigate alternatives to PCA.
\label{sec:embd_stats}
\begin{table}[h]
  \caption{
  Comparison of performance of three types of embeddings extracted from  ResNet34 with temporal stride 4 and std pooling mechanism. } 
  \label{tab:results:embd_stats}
  \centering
   \begin{tabular}{ c l  c c  c c }

    \toprule
    && \multicolumn{2}{c}{\textbf{cos}}&\multicolumn{2}{c}{\textbf{mix PLDA}}\\
     & \textbf{} & 
      \textbf{min\_C}  & \textbf{EER (\%)}& \textbf{min\_C}  & \textbf{EER (\%)}  \\
    \midrule
1 &embd & 0.662& 15.60	&0.473 & 8.38 \\ 
    2 & stats+PCA &0.911&27.57&0.527& 9.51   \\ 
    3&cl-embd&0.612&12.22	&0.475&8.35\\

    \bottomrule
  \end{tabular}
  \vspace{-4mm}
\end{table}

As both types of embedding extractors that we use here were trained on exactly the same dataset and the main difference between them is only the size of the network, when fusing two systems (either using score level fusion or the approach described in appendix~\ref{logit_fusion}) we do not observe any performance improvements compared to using just the bigger ResNet152 model. Thus, we do not report any fusion results here. 

\subsection{Fine-tuning the extractor on long segments}
\label{sec:ft}
Previously (see e.g.~\cite{garcia2020magneto}) and in system descriptions of several top performing teams of NIST SRE 2021 (e.g. in~\cite{stc_sysdescr}), it was noted that fine-tuning (FT) a  pretrained embedding extractor on long segments (10-20s vs. 2-4s) results in a major performance gain. In order to verify this finding, we perform the following experiment: we fine-tune our two embedding extractors on 10s long segments. The embeddings extracted from the fine-tuned models are scored with the approach described in Section~\ref{sec:backend_theory}.

The results of this experiment are presented in Table~\ref{tab:results:ft}. It is seen that for both extractors fine-tuning procedure provides a considerable performance improvement. Also, we can note that the effect of the fine-tuning does not depend on the back-end approach: in previous works, either cosine scoring or a single PLDA model were used, while we observe similar performance improvements with a mixture of PLDA models. 

\begin{table}[th]
  \caption{
 Effect of fine-tuning embedding extractors on 10s long segments on speaker verification performance. Two types of networks are used: ResNet34 with std pooling and ResNet152 with mean and std pooling. For both networks, temporal stride is set to 4. The back-end in all cases is a mixture of 3 language-dependent PLDA models.} 
  \label{tab:results:ft}
  \centering
 \begin{tabular}{ c l  c c  c c}

    \toprule
     & \multirow{2}{*}{\textbf{FT}} & 
     \multicolumn{2}{c}{\textbf{ResNet34}} &\multicolumn{2}{c}{\textbf{ResNet152}}  \\
     & & \textbf{min\_C}  & \textbf{EER (\%)} & \textbf{min\_C} & \textbf{EER (\%)} \\
    \midrule
1 & no & 0.473 & 8.38 & 0.407 & 6.28 \\ 
    2 & yes & 0.432& 7.38 &0.380 & 5.70  \\ 

    \bottomrule
  \end{tabular}
  \vspace{-4mm}
\end{table}

\subsection{Back-end processing}
\label{sec:backend}
In this section, we investigate the impact of individual steps of the back-end recipe described in Section~\ref{sec:backend_theory} compared to a standard back-end approach: a single PLDA model trained on the embeddings that were centered, their dimensionality reduced from 256 to 200 by LDA and then length-normalized. Table~\ref{tab:results:backend} presents gradual improvements in the verification performance with step-by-step transformation of the baseline back-end into the back-end described in Section~\ref{sec:backend_theory}. The same experiments were performed for embeddings extracted with ResNet34 and ResNet152 models with temporal strides set to [1,2,1,2]. 

\begin{figure}[tb]
\centering
\includegraphics[width=1.0\linewidth]{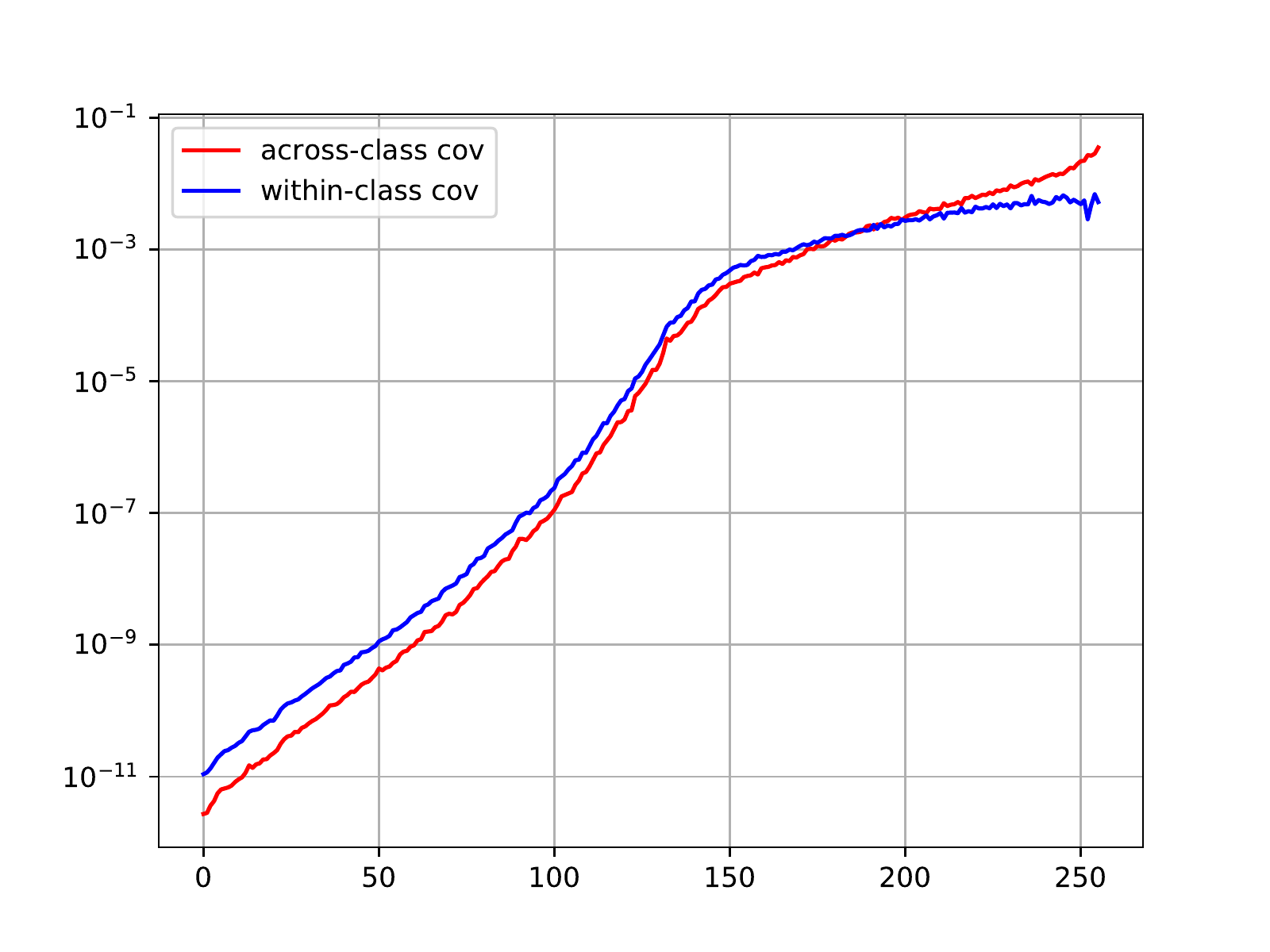}
\vspace{-4mm}
\caption{Diagonal across- and within-class covariances of the back-end training data. Prior to plotting the graphs, the embeddings were projected onto eigenvectors of the total covariance matrix i.e. total covariance is diagonal with elements sorted by the index on horizontal axis. }
\label{fig:eigs_wc_ac}
\vspace{-4mm}
\end{figure}
From the results, the largest improvement was observed when reducing the dimensionality of the embedding after LDA from 200 to 100 (lines 1 and 2 of Table~\ref{tab:results:backend}). Figure~\ref{fig:eigs_wc_ac} gives an insight into this phenomenon. It displays eigenvalues of the across- and within-class covariance matrices estimated on the PLDA training data. The eigenvalues were computed after projecting the embeddings with eigenvectors of the total covariance matrix i.e. the elements of total covariance monotonically increase from left to right. Figure~\ref{fig:eigs_wc_ac} shows the plots for ResNet34 embedding extractor, similar pattern is observed for ResNet152. By inspecting the graphs, we notice that roughly half of the embedding dimensions have very little variability. When such directions are selected by LDA, their (possibly noisy) variability is amplified leading to performance degradation.

Introducing a mixture of language-dependent PLDA models instead of a single one (lines 2 and 3 of the table) brings another considerable improvement. Finally, removing the direction corresponding to the highest gender variability with NAP results in an additional performance gain. However, compared to the previous two modifications the improvement from using NAP is rather small.
\begin{table*}
  \caption{
  Effect of the individual steps used for back-end training. The results are shown for ResNet34 with std pooling and ResNet152 with mean and std pooling. Temporal stride for both extractors is set to 4.} 
  \label{tab:results:backend}
  \centering
  \begin{tabular}{ c l  c c  c c}

    \toprule
     & \multirow{2}{*}{\textbf{Back-end}} & 
     \multicolumn{2}{c}{\textbf{ResNet34}} &\multicolumn{2}{c}{\textbf{ResNet152}}  \\
     & & \textbf{min\_C}  & \textbf{EER (\%)} & \textbf{min\_C} & \textbf{EER (\%)} \\
    \midrule

1 & LDA200, LN, PLDA & 0.584 & 11.37 &0.569&9.90\\ 
    2 & LDA100, LN, PLDA & 0.515 & 10.16&0.450& 7.49  \\ 
3 & LDA100, LN, 3PLDAs & 0.481 & 8.76 &0.419&6.46  \\ 
4 & NAP gender, LDA100, LN, 3PLDAs & 0.473 & 8.38 &0.407&6.28  \\ 
    \bottomrule
  \end{tabular}
\end{table*}

\subsection{Alternative back-end approaches}
Here, we want to compare the back-end strategy presented in Section~\ref{sec:backend_theory} with some other approaches that were shown to be effective in NIST SRE21. Namely, we have noticed that many teams used cosine similarity scoring instead of PLDA back-end; in~\cite{stc_sysdescr}, it was shown that it is beneficial to use score and channel normalization when doing cosine similarity scoring of the embeddings. Also, as mentioned in Section~\ref{sec:embd_stats}, cl-embeddings (logits from the embedding extractor network) in combination with cosine scoring were shown to provide competitive speaker verification results. In order to show the effectiveness of the mentioned approaches we use three scoring strategies: cosine scoring on the embeddings and on the cl-embeddings and our approach with NAP, LDA and a mixture of 3 PLDA models. For each of these strategies, we report the results with and without score, channel, and combination of both normalization.

The cohort for score normalization is constructed from NIST CTS Superset by averaging the embeddings on per-speaker basis i.e. the cohort contains one embedding per speaker from the PLDA training set resulting in approximately 7k embeddings. We use 400 highest scores of the enrollment and test segments against the cohort for score normalization.

Channel normalization is a calibration-like technique that shifts and scales evaluation scores based on the channel type of the enrollment-test pair. For each of four types of trials (mic-mic, tel-tel, mic-tel, tel-mic), we evaluate the mean and standard deviation of the scores from NIST SRE21 development set. These parameters are then used to shift and scale evaluation scores for the matching trial type.  If both score and channel normalization are used, we perform score normalization first. When scoring cl-embeddings, rather than computing cosine distance scores directly we follow the approach described in Section~\ref{sec:embd_stats} allowing for efficient score estimation. Note that the channel is assumed given in SRE21. In real systems, one should estimate it using a channel classifier.

We report the results of the aforementioned approaches on ResNet34 and ResNet152 embeddings with and without fine-tuning on the long speech segments.
The results are shown in Table~\ref{tab:results:alt_appr}.
Several conclusions could be made looking at the results. First, as expected, we notice that for any kind of back-end or score normalization, the bigger ResNet152 provides lower error rates than the smaller ResNet34. Second, our results indicate that channel normalization (and its combination with score normalization) improves the performance only for cosine distance scoring and for the models that were fine-tuned on long speech segments. For the models without fine-tuning, we observe, that typically channel normalization results in the improvements in terms of EER and performance drop in terms of min\_C. Also, we verify the observation of~\cite{stc_sysdescr} that cosine scoring of cl-embeddings outperforms cosine scoring of conventional low-dimensional embeddings.  However, comparing the results of the mixture of PLDA models with the simple cosine scoring of cl-embeddings, we note that the more elaborated approach provides considerably better results for all embedding extractors. Thus, we conclude that even though training separate a back-end model adds additional complexity to the pipeline, it pays off in terms of the system performance.
\begin{table*}
  \caption{Comparison of speaker verification performance of different scoring approaches. Two types of embedding extractors are used (ResNet34 with standard deviation pooling and ResNet152 with mean and std pooling, cumulative temporal stride 4 is used for both networks). For both extractors, we present the results with and without fine-tuning on 10s segments.} 
  \label{tab:results:alt_appr}
  \centering
  \begin{tabular}{  c c c  c c  c c c c c c}

    \toprule
      \multirow{3}{*}{\textbf{Back-end}} &\multirow{3}{*}{\textbf{S-norm}}&\multirow{3}{*}{\textbf{Ch-norm}} &
     \multicolumn{4}{c}{\textbf{ResNet34}} &\multicolumn{4}{c}{\textbf{ResNet152}}  \\
      \textbf{} & &&
     \multicolumn{2}{c}{\textbf{no FT}} &\multicolumn{2}{c}{\textbf{FT}} &\multicolumn{2}{c}{\textbf{no FT}} &\multicolumn{2}{c}{\textbf{FT}}\\
      &&& \textbf{min\_C}  & \textbf{EER (\%)} & \textbf{min\_C} & \textbf{EER (\%)} & \textbf{min\_C}  & \textbf{EER (\%)} & \textbf{min\_C} & \textbf{EER (\%)}\\
    \midrule

  \multirow{4}{*}{cos} & no&no &0.662& 15.6		&0.554&10.89&	0.589&		12.22&	0.49&	9.18\\ 
    & yes&no & 0.65&15.25&	0.58&	10.52&0.606&		11.91&	0.525&	8.68\\ 
    & no&yes & 0.706&15.35&	0.552&	10.31&0.62&		12.6&	0.498&	8.77\\ 
    & yes&yes & 0.664&14.89&	0.507&	9.9&0.615&		12.28&	0.478&	8.51\\ 
  \midrule
    \multirow{4}{*}{cos cl-embd} & no&no & 0.612&12.22&	0.534&	9.61&0.528&		9.76&	0.456&	7.72\\ 
    & yes&no & 0.612&11.71&	0.57&	9.12&0.538&		9.23&	0.478&	7.3\\ 
    & no&yes & 0.69&12.42&		0.548&8.83&0.574&		9.85&	0.458&	7.39\\ 
    & yes&yes & 0.623&11.52&	0.49&	8.04&0.546&		9.38&	0.429&	6.83\\ 
  \midrule
     & no&no & 0.473&8.38&	0.432&	7.38&0.407&	6.28&	0.38&	5.7\\ 
   NAP, LDA, & yes&no & 0.476&8.42&	0.426&	7.48&0.406&	6.27&	0.372&	5.77\\ 
   \multirow{2}{*}{ mix 3 PLDA} & no&yes & 0.592&8.78&	0.526&	7.49&0.493&	6.71&	0.433&	5.83\\ 
    & yes&yes &0.543&8.89&	0.47&	7.56&0.44&	6.59&	0.38&	5.73\\ 
    \bottomrule
  \end{tabular}
\end{table*}

\subsection{Nuisance variability suppression}
\label{sec:nap}
As was noticed in Section~\ref{sec:backend}, using NAP to remove the direction corresponding to the highest gender variability from the data resulted in a slight performance improvement. A reasonable question then is whether the gender dimension was the one to remove or there are other sources of unwanted variability in the embedding distribution that we have to deal with. To analyze this question we performed the following experiment: we train several PLDA models, each of them is trained on the same embeddings. What differs between these models is the pre-processing performed on the embeddings: we either do just centering, LDA and LN, or, prior to centering, we remove a few directions from the data corresponding to various sources of variability. Namely, we remove 2 directions corresponding to the largest PCA eigenvalues i.e. the directions of the highest variability of the data (we expect that these should correspond to gender variability), alternatively, we apply NAP to remove gender, language, or dataset (individual subsets of CTS superset) variability. In all of these experiments, we use a single PLDA to isolate the effect of pre-processing.

The results of this experiment are displayed in Table~\ref{tab:results:nap}, where lines correspond to the different pre-processing steps described above. The table shows the results for ResNet34 that was or was not fine-tuned on long training segments. As results suggest, when the extractor is not fine-tuned on long segments any kind of variability compensation does not provide any significant improvement in terms of min\_C, while removing the dimensions corresponding to either gender or language results in some improvement in terms of EER. However, when the network was fine-tuned, there is a clear performance gain from using gender NAP compared to any other variability compensation approaches. We do not have a clear explanation of this phenomenon, it has to be investigated further.

\begin{table}
  \caption{Comparison of various nuisance variability suppression approaches. The embedding extractor is ResNet34 with std pooling and cumulative temporal stride set to 4. FT denotes fine-tuning on 10s speech segments.} 
  \label{tab:results:nap}
  \centering
   \begin{tabular}{ c l  c c  c c}

    \toprule
     & \multirow{2}{*}{\textbf{pre-process}} & 
     \multicolumn{2}{c}{\textbf{no FT}} &\multicolumn{2}{c}{\textbf{FT}}  \\
     & & \textbf{min\_C}  & \textbf{EER (\%)} & \textbf{min\_C} & \textbf{EER (\%)} \\
    \midrule
1 & - & 0.515 & 10.16 & 0.499& 10.02 \\ 
    2 & PCA top 2 & 0.519 & 10.07 & 0.504&9.94  \\ 
    3  & NAP gender & 0.511 & 9.71 & 0.463&8.51\\ 
    4  & NAP lang & 0.514 & 9.90 & 0.493&9.48\\ 
    5  & NAP db & 0.517 & 10.12 & 0.490&9.58\\ 
    \bottomrule
  \end{tabular}
\end{table}

\section{Conclusion}
In this paper, we proposed a set of methods to improve speaker recognition performance on a challenging NIST SRE 2021. The setup is characterized by language mismatch between trials and limited or no in-domain data for training or fine-tuning models. Our contributions are related to certain architectural improvements of the extractor (such as the use of reduced temporal stride and the use of alternative pooling methods) as well as to the back-end of the system (such as the use of a PLDA mixture and NAP). We also evaluated our methods against a strong novel approach that uses cosine similarity with score and channel normalization, and logits instead of embeddings, for which we demonstrated a more efficient implementation with embedding dimensional speaker representations. The experiments showed that our proposed approach outperforms the baseline, while it does not make use of any channel (oracle or real) classifier, justifying the use of a more probabilistic back-end in setups similar to SRE21.   

\section{Acknowledgements}
This project has received funding from the European Union’s Horizon 2020 research and innovation programme under grant agreements No. 101007666 / ESPERANTO / H2020-MSCA-RISE-2020 and No. 833635 / ROXANNE.

\bibliographystyle{IEEEbib}
\bibliography{Odyssey2022_BibEntries}

\begin{thebibliography}{10}

\bibitem{DehakN_TASLP:2010}
N.~Dehak, P.~Kenny, R.~Dehak, P.~Dumouchel, and P.~Ouellet,
\newblock ``Front-{E}nd {F}actor {A}nalysis {F}or {S}peaker {V}erification,''
\newblock {\em IEEE Transactions on Audio, Speech, and Language Processing},
  vol. 19, no. 4, pp. 788--798, May 2011.

\bibitem{prince:iccv:2007}
S.~J.~D. Prince and J.~H. Elder,
\newblock ``{P}robabilistic linear discriminant analysis for inferences about
  identity,''
\newblock in {\em Proc. International Conference on Computer Vision (ICCV)},
  Rio de Janeiro, Brazil, 2007.

\bibitem{lozano_odyssey_2016}
A.~Lozano-Diez, A.~Silnova, P.~Mat{\v{e}}jka, O.~Glembek, O.~Plchot,
  J.~Pe{\v{s}}{\'{a}}n, L.~Burget, and J.~Gonzalez-Rodriguez,
\newblock ``{A}nalysis and {O}ptimization of {B}ottleneck {F}eatures for
  {S}peaker {R}ecognition,''
\newblock in {\em Proceedings of Odyssey 2016}. 2016, vol. 2016, pp. 352--357,
  International Speech Communication Association.

\bibitem{Matejka:ICASSP:2016}
Pavel Mat{\v{e}}jka, Ond{\v{r}}ej Glembek, Ond{\v{r}}ej Novotn{\'{y}},
  Old{\v{r}}ich Plchot, Franti{\v{s}}ek Gr{\'{e}}zl, Luk{\'{a}}{\v{s}} Burget,
  and Jan {\v{C}}ernock{\'{y}},
\newblock ``{A}nalysis {O}f {DNN} {A}pproaches {T}o {S}peaker
  {I}dentification,''
\newblock in {\em Proceedings of ICASSP}. 2016, pp. 5100--5104, IEEE Signal
  Processing Society.

\bibitem{jung2019rawnet}
Jee-weon Jung, Hee-Soo Heo, Ju-ho Kim, Hye-jin Shim, and Ha-Jin Yu,
\newblock ``Rawnet: Advanced end-to-end deep neural network using raw waveforms
  for text-independent speaker verification.,''
\newblock {\em Proc. Interspeech 2019}, pp. 1268--1272, 2019.

\bibitem{jung2020improved}
Jee-weon Jung, Seung-bin Kim, Hye-jin Shim, Ju-ho Kim, and Ha-Jin Yu,
\newblock ``Improved rawnet with feature map scaling for text-independent
  speaker verification using raw waveforms,''
\newblock {\em Proc. Interspeech 2020}, pp. 1496--1500, 2020.

\bibitem{lin2020wav2spk}
Weiwei Lin and Man-Wai Mak,
\newblock ``Wav2spk: A simple dnn architecture for learning speaker embeddings
  from waveforms,''
\newblock {\em Proc. Interspeech 2020}, pp. 3211--3215, 2020.

\bibitem{Lei_icassp_2014}
Y.~Lei, N.~Scheffer, L.~Ferrer, and M.~McLaren,
\newblock ``A novel scheme for speaker recognition using a phonetically-aware
  deep neural network,''
\newblock in {\em Proceedings of ICASSP}, May 2014, pp. 1695--1699.

\bibitem{kennydeep}
P~Kenny, V~Gupta, T~Stafylakis, P~Ouellet, and J~Alam,
\newblock ``Deep neural networks for extracting baum-welch statistics for
  speaker recognition,''
\newblock in {\em Proc. Odyssey}, 2014.

\bibitem{Novoselov_interspeech_2015}
S.~Novoselov, T.~Pekhovsky, O.~Kudashev, V.~S. Mendelev, and A.~Prudnikov,
\newblock ``{N}on-linear {PLDA} for i-vector speaker verification,''
\newblock in {\em Proceedings of ICASSP}, Sept 2015, pp. 214--218.

\bibitem{Bhattacharya_SLT16}
G.~Bhattacharya, J.~Alam, P.~Kenny, and V.~Gupta,
\newblock ``Modelling speaker and channel variability using deep neural
  networks for robust speaker verification,''
\newblock in {\em 2016 {IEEE} Spoken Language Technology Workshop, {SLT} 2016,
  San Diego, CA, USA, December 13-16}, 2016.

\bibitem{Ghahabi_icassp_2014}
O.~Ghahabi and J.~Hernando,
\newblock ``Deep belief networks for i-vector based speaker recognition,''
\newblock in {\em Proceedings of ICASSP}, May 2014, pp. 1700--1704.

\bibitem{heighold_icassp_2016}
G.~Heigold, I.~Moreno, S.~Bengio, and N.~Shazeer,
\newblock ``End-to-end text-dependent speaker verification,''
\newblock in {\em Proceedings of ICASSP}, March 2016, pp. 5115--5119.

\bibitem{zhang_slt_2016}
S.~X. Zhang, Z.~Chen, Y.~Zhao, J.~Li, and Y.~Gong,
\newblock ``{E}nd-to-{E}nd attention based text-dependent speaker
  verification,''
\newblock in {\em 2016 IEEE Spoken Language Technology Workshop (SLT)}, Dec
  2016, pp. 171--178.

\bibitem{sreEndEnd:Snyder2016}
D.~Snyder, P.~Ghahremani, D.~Povey, D.~Garcia-Romero, Y.~Carmiel, and
  S.~Khudanpur,
\newblock ``Deep neural network-based speaker embeddings for end-to-end speaker
  verification,''
\newblock in {\em 2016 IEEE Spoken Language Technology Workshop (SLT)}, Dec
  2016, pp. 165--170.

\bibitem{rohdin:icassp:2018}
Johan Rohdin, Anna Silnova, Mireia Diez, Old{\v{r}}ich Plchot, Pavel
  Mat\v{e}jka, and Luk{\'{a}}{\v{s}} Burget,
\newblock ``{E}nd-to-end {DNN} based speaker recognition inspired by i-vector
  and {PLDA},''
\newblock in {\em Proceedings of ICASSP}. 2018, IEEE Signal Processing Society.

\bibitem{bhattacharya2016deep}
Gautam Bhattacharya, Jahangir Alam, Themos Stafylakis, and Patrick Kenny,
\newblock ``Deep neural network based text-dependent speaker recognition:
  Preliminary results,''
\newblock in {\em Proc. Odyssey}, 2016.

\bibitem{Variani_icassp_2014}
E.~Variani, X.~Lei, E.~McDermott, I.~L. Moreno, and J.~Gonzalez-Dominguez,
\newblock ``Deep neural networks for small footprint text-dependent speaker
  verification,''
\newblock in {\em Proceedings of ICASSP}, May 2014, pp. 4052--4056.

\bibitem{Bhattacharaya_interspeech_2017}
G.~Bhattacharya, J.~Alam, and P.~Kenny,
\newblock ``{D}eep {S}peaker {E}mbeddings for {S}hort-{D}uration {S}peaker
  {V}erification,''
\newblock in {\em Interspeech 2017}, 08 2017, pp. 1517--1521.

\bibitem{xvec:Snyder2016}
David Snyder, Daniel Garcia-Romero, Daniel Povey, and Sanjeev Khudanpur,
\newblock ``{D}eep {N}eural {N}etwork {E}mbeddings for {T}ext-{I}ndependent
  {S}peaker {V}erification,''
\newblock {\em Proc. Interspeech 2017}, pp. 999--1003, 2017.

\bibitem{stafylakis2019self}
Themos Stafylakis, Johan Rohdin, Old{\v{r}}ich Plchot, Petr Mizera, and
  Luk{\'a}{\v{s}} Burget,
\newblock ``Self-supervised speaker embeddings,''
\newblock in {\em Proc. Interspeech 2019}, 2019, pp. 2863--2867.

\bibitem{he2016deep}
Kaiming He, Xiangyu Zhang, Shaoqing Ren, and Jian Sun,
\newblock ``Deep residual learning for image recognition,''
\newblock in {\em Proceedings of the IEEE conference on computer vision and
  pattern recognition}, 2016, pp. 770--778.

\bibitem{Zeinali_VOXCELEB_2019}
Hossein Zeinali, Shuai Wang, Anna Silnova, Pavel Mat\v{e}jka, and Old\v{r}ich
  Plchot,
\newblock ``But system description to voxceleb speaker recognition challenge
  2019,''
\newblock in {\em Proceedings of The VoxCeleb Challange Workshop 2019}, 2019,
  pp. 1--4.

\bibitem{ABC_SRE_CTS_2019}
Jahangir Alam, Gilles Boulianne, Ond\v{r}ej Glembek, Alicia~D\'{i}ez Lozano,
  Pavel Mat\v{e}jka, Petr Mizera, Joao Monteiro, Ladislav Mo\v{s}ner,
  Ond\v{r}ej Novotn\'{y}, Old\v{r}ich Plchot, A.~Johan Rohdin, Anna Silnova,
  Josef Slav\'{i}\v{c}ek, Themos Stafylakis, Shuai Wang, and Hossein Zeinali,
\newblock ``{ABC NIST SRE 2019 CTS System Description},''
\newblock in {\em Proceedings of NIST}. 2019, pp. 1--6, National Institute of
  Standards and Technology.

\bibitem{zhou2021resnext}
Tianyan Zhou, Yong Zhao, and Jian Wu,
\newblock ``Resnext and res2net structures for speaker verification,''
\newblock in {\em 2021 IEEE Spoken Language Technology Workshop (SLT)}. IEEE,
  2021, pp. 301--307.

\bibitem{wang2021revisiting}
Shuai Wang, Yexin Yang, Yanmin Qian, and Kai Yu,
\newblock ``Revisiting the statistics pooling layer in deep speaker embedding
  learning,''
\newblock in {\em 2021 12th International Symposium on Chinese Spoken Language
  Processing (ISCSLP)}. IEEE, 2021, pp. 1--5.

\bibitem{stafylakis2021speaker}
Themos Stafylakis, Johan Rohdin, and Luk{\'a}{\v{s}} Burget,
\newblock ``Speaker embeddings by modeling channel-wise correlations,''
\newblock in {\em Proc. Interspeech 2021}, 2021, pp. 501--505.

\bibitem{snyder2018Xvectors}
D.~{Snyder}, D.~{Garcia-Romero}, G.~{Sell}, D.~{Povey}, and S.~{Khudanpur},
\newblock ``X-vectors: Robust dnn embeddings for speaker recognition,''
\newblock in {\em Proceedings of ICASSP}, 2018, pp. 5329--5333.

\bibitem{stc_sysdescr}
A.~Avdeeva, A.~Gusev, I.~Korsunov, A.~Kozlov, G.~Lavrentyeva, S.~Novoselov,
  T.~Pekhovsky, A.~Shulipa, A.~Vinogradova, V.~Volokhov, et~al.,
\newblock ``{STC speaker recognition systems for the NIST SRE 2021},''
\newblock {\em arXiv preprint arXiv:2111.02298}, 2021.

\bibitem{solomonoff2005}
A.~Solomonoff, W.M. Campbell, and I.~Boardman,
\newblock ``Advances in channel compensation for svm speaker recognition,''
\newblock in {\em Proceedings of ICASSP}, 2005, vol.~1, pp. I/629--I/632 Vol.
  1.

\bibitem{aronowitz2014inter}
Hagai Aronowitz,
\newblock ``Inter dataset variability compensation for speaker recognition,''
\newblock in {\em Proceedings of ICASSP}. IEEE, 2014, pp. 4002--4006.

\bibitem{senoussaoui2011mixture}
Mohammed Senoussaoui, Patrick Kenny, Niko Br{\"u}mmer, Edward~de Villiers, and
  Pierre Dumouchel,
\newblock ``Mixture of plda models in i-vector space for gender-independent
  speaker recognition,''
\newblock in {\em Twelfth Annual Conference of the International Speech
  Communication Association}, 2011.

\bibitem{garcia2012multicondition}
D.~Garcia-Romero, X.~Zhou, and C.~Y. Espy-Wilson,
\newblock ``Multicondition training of gaussian plda models in i-vector space
  for noise and reverberation robust speaker recognition,''
\newblock in {\em Proceedings of ICASSP}. IEEE, 2012, pp. 4257--4260.

\bibitem{BrummerSimPLDA}
Niko Brummer,
\newblock ``{EM for Simple PLDA},'' 2010,
\newblock Technical report.

\bibitem{sre_cts_superset}
Omid Sadjadi,
\newblock ``{NIST SRE CTS Superset: A large-scale dataset for telephony speaker
  recognition},'' 2021-08-16 04:08:00 2021.

\bibitem{Nagrani19}
Arsha Nagrani, Joon~Son Chung, Weidi Xie, and Andrew Zisserman,
\newblock ``Voxceleb: Large-scale speaker verification in the wild,''
\newblock {\em Computer Speech and Language}, 2019.

\bibitem{chung2018voxceleb2}
Joon~Son Chung, Arsha Nagrani, and Andrew Zisserman,
\newblock ``Voxceleb2: Deep speaker recognition,''
\newblock {\em arXiv preprint arXiv:1806.05622}, 2018.

\bibitem{snyder2015musan}
David Snyder, Guoguo Chen, and Daniel Povey,
\newblock ``Musan: A music, speech, and noise corpus,''
\newblock {\em arXiv preprint arXiv:1510.08484}, 2015.

\bibitem{Sadjadi:2021}
O.~Sadjadi, C.~Greenberg, E.~Singer, L.~Mason, and D.~Reynolds,
\newblock ``{NIST 2021 Speaker Recognition Evaluation Plan, NIST SRE},'' 2021.

\bibitem{garcia2020magneto}
D.~Garcia-Romero, G.~Sell, and A.~Mccree,
\newblock ``Magneto: X-vector magnitude estimation network plus offset for
  improved speaker recognition,''
\newblock in {\em Proc. Odyssey 2020 the speaker and language recognition
  workshop}, 2020.

\bibitem{deng2019arcface}
Jiankang Deng, Jia Guo, Niannan Xue, and Stefanos Zafeiriou,
\newblock ``Arcface: Additive angular margin loss for deep face recognition,''
\newblock in {\em Proceedings of the IEEE/CVF Conference on Computer Vision and
  Pattern Recognition}, 2019, pp. 4690--4699.

\end{thebibliography}

\appendix
\section{Architectural details, optimizers and training strategies}
\subsection{ResNet34}
\label{sec:resnet34}
The network is composed on 34 convolutional layers with residual connections. All convolutional kernels are 3×3, the number of channels is (64,128,256,256) and the first convolutional layer also outputs 64 channels. The number of convolutional layers per block is (3, 4, 6, 3). The input features are 64-dimensional fbanks, extracted from 8kHz audio files, and the training segments contain 350 frames.

The networks are trained using multi-speaker classification and with Additive Angular Margin loss with 30 and 0.3 scale and margin, respectively \cite{deng2019arcface}. As optimizer we use stochastic gradient descent with momentum equal to 0.9. The minibatch size is 256, however to fit it in a single GPU we split the minibatch into 16 “microbatches” of 16 examples each and use gradient accumulation. The initial learning rate is 0.2 which we divide by 2 when the loss does not improve for more than 3000 model updates in the held-out set. When the network is fine-tuned on long speech segments, we keep the margin of AAM softmax at a value of 0.3. The initial learning rate is set to 0.01 and gradually decreased upon reaching a plateau of cross-validation loss until convergence.

\subsection{ResNet152} 
\label{sec:resnet152}
The ResNet152 embedding extractor comprises 152 convolutional layers. The stages of the network consist of 3, 8, 36, and 3 bottleneck blocks with a pre-activation structure and use 64, 128, 256, and 256 channels, respectively. Per-frame representations are aggregated with a traditional mean and standard deviation pooling. The resulting vectors are projected to 256-dimensional embeddings. The model requires 64-dimensional fbank features and is trained to optimize the AAM loss.

\section{Efficient fusion in the logit domain} 
\label{logit_fusion}
The fusion method in the logit space of two or more extractors trained with the same set of speakers (suggested in~\cite{stc_sysdescr}) can also be performed in a low-dimensional space. More specifically, assume two networks with embedding dimension equal to $d_1$ and $d_2$. The motivation of~\cite{stc_sysdescr} is to fuse the two systems using the weighted average logit vectors $\lvec_f = w_1\lvec_1 + w_2 \lvec_2$, assuming that the number of training speakers and their order (as encoded in $\lvec$ via $\Kmat$) is the same in both systems. To derive the cosine similarity in the low-dimensional space ($d_f = d_1 + d_2$) we may define the embedding $\rvec_f = [w_1 \rvec'_1, w_2 \rvec'_2]'$ of length $d_f$. Using linear algebra, one may verify that the cosine similarity between two fused logit vectors $\lvec_{f,e}$ and $\lvec_{f,t}$ is identical to the cosine similarity between $\Mmat_f\rvec_{f,e}$ and $\Mmat_f\rvec_{f,t}$. $\Mmat_f$ is the upper triangular matrix of the Cholesky decomposition of the between-speaker covariance in the $d_f$-dimensional space, i.e. $\Mmat'_f \Mmat_f = \Kmat'_f \Kmat_f$, where $\Kmat_f = [\Kmat_1,\Kmat_2]$ is the matrix of concatenated projection matrices of shape $N_s \times d_f$. Therefore, logit-domain fusion can be performed in the $d_f$-dimensional space, by projecting the concatenated embeddings $\rvec_f$ onto $\Mmat_f$. 

%

\end{document}